\documentclass[10pt, journal, doublecolumn]{IEEEtran}
\usepackage{amsfonts}
\usepackage{mathrsfs}
\usepackage{bbm}
\usepackage{bm}
\usepackage{cite}      

\usepackage{graphicx}  

\usepackage{stfloats}  
\usepackage{amsmath}   
\usepackage{amssymb}
\usepackage{amsfonts}
\usepackage{flushend,cuted}
\newtheorem{rem}{Remark}
\begin{document}
\newcommand{\e}[1]{\boldsymbol{#1}}

\title{Deep Learning Based Pilot Design for Multi-user Distributed Massive MIMO Systems}

\IEEEoverridecommandlockouts
\author{Jun~Xu,~\IEEEmembership{Student Member,~IEEE},
        Pengcheng~Zhu,~\IEEEmembership{Member,~IEEE},
        Jiamin~Li,
        and~Xiaohu~You,~\IEEEmembership{Fellow,~IEEE}\\
\thanks{This work was supported in part by the Natural Science Foundation of Jiangsu Province (Grant No.
20180011), and in part by the National Natural Science Foundation of China (Grant Nos. 61571120 and 61501113).}
\thanks{J. Xu, P. Zhu, J. Li, and X. You are all with National Mobile Communications Research Laboratory, Southeast University, Nanjing, China ( emails: \{xujunseu, p.zhu, lijiamin, xhyu\}@seu.edu.cn).}
}
\maketitle


\begin{abstract}
This letter proposes a deep learning based pilot design scheme to minimize the sum mean square error (MSE) of channel estimation for multi-user distributed massive multiple-input multiple-output (MIMO) systems. The pilot signal of each user is expressed as a weighted superposition of orthonormal
pilot sequence basis, where the power assigned to each pilot sequence is the corresponding weight. A multi-layer fully connected deep neural network (DNN) is designed to optimize the power allocated to each pilot sequence to minimize the sum MSE, which takes the channel large-scale fading coefficients as input and outputs the pilot power allocation vector. The loss function of the DNN is defined as the sum MSE, and we leverage the unsupervised learning strategy to train the DNN. Simulation results show that the proposed scheme achieves better sum MSE performance than other methods with low complexity.
\end{abstract}


\section{Introduction}
Distributed massive multiple-input multiple-output (MIMO) systems have emerged as a promising antenna topology in 5G wireless communication, due to
its advantage in improving spectral and energy efficiency, as well as lowering cutoff rate\cite{XHYou10}.
Channel state information (CSI) is essential to most of signal processing in wireless communication, such
as signal detection and beamforming. It is usually acquired by transmitting pilot sequences. However, due to limited time and
frequency resources, some users have to reuse the same orthogonal pilot sequences, which leads to pilot contamination. Hence, more and more people devote themselves to the research of alleviating pilot contamination in distributed massive MIMO systems.

To the best of our knowledge, the most effective method to alleviate pilot contamination is optimizing pilot assignment schemes among users. The optimal pilot assignment scheme can be obtained by exhaustively searching all possible cases. However, the exponential complexity of exhaustive search makes it impractical. To achieve a tradeoff between the performance and complexity, some heuristic algorithms were proposed to allocate pilot sequences based on the channel covariance matrix\cite{HYin13, LYou15}. Unfortunately, most of heuristic algorithms can not achieve a satisfying tradeoff in a practical system. Recently, \cite{TVan} proposed a novel pilot design where the pilot signals are a weighted superposition of orthonormal pilot basis vectors, and treated the associated pilot power coefficients as continuous variables, which can be optimized by utilizing optimization theory. However, the complexity of above-mentioned methods are still very high.

With the rapid development of deep learning, it has gradually become a promising tool in solving difficult wireless communication problems due to its excellent performance and low complexity, such as resource allocation\cite{liang2018towards}, channel decoding\cite{FLiangJSP} and channel estimation \cite{KKim, DNeumann}. \cite{KKim} designed a supervised learning based pilot assignment scheme for a massive MIMO system, in which the network is trained with the optimal pilot assignment results of exhaustive search serving as the ground truth. However, this method is applicable only when the search space is small because of the exponential complexity of exhaustive search.

In this letter, we propose a pilot design problem, which optimizes the power allocated to each pilot sequence for each user, to minimize the sum mean square error (MSE) of channel estimation for a distributed massive MIMO system. It is a NP-hard problem, and we propose an unsupervised learning based pilot power allocation scheme to solve it. Specifically, a deep neural network (DNN) is designed to train the mapping from the input (channel large-scale fading coefficients) to the output (pilot power allocation vector). In the training process, we select the sum MSE as the loss function, and continually train the network parameters to adjust the output, further minimizing the sum MSE. Compared to the supervised learning scheme adopted in \cite{KKim}, the ground truth is not necessary in unsupervised learning, thus the proposed pilot power allocation scheme can be applied to the case with large number of users. Simulation results show that the proposed scheme achieves better performance than traditional pilot assignment scheme with lower complexity.

The notations used in this paper are conformed to the following
convention. Boldface letters stand for matrices (upper case) or vectors (lower case). The transpose and conjugate transpose are denoted by ${\left(  \cdot  \right)^ \textrm{T} }$ and ${\left(  \cdot  \right)^ \textrm{H} }$ respectively. ${{\bf{I}}_M}$ stands for the $M\times M$ identity
matrix, and $\mathcal{\mathcal{N}_{\rm{c}}}(\mu,\sigma^2)$ denotes the
circularly symmetric complex Gaussian distribution with mean $\mu$
and variance $\sigma^2$.


\section{System Model}
This paper considers a multi-user (MU) distributed massive MIMO system. There are $M$ remote antenna units (RAUs) equipped with $N$
antennas and $K$ single-antenna users that share the same bandwidth in the cell. We let ${\cal M} = \left\{ {1, \ldots M} \right\}$ and ${\cal K} = \left\{ {1, \ldots K} \right\}$ represent the set of RAUs and users respectively.

\subsection{Channel Model}
This paper considers a block flat-fading channel, and the channel vector from user $k$ to all RAUs can be modeled as\cite{cao2018uplink}
\begin{equation}
    {{\mathbf{g}}_k}{\rm{ = }}{\bf{\Lambda }}_k^{1/2}{{\mathbf{h}}_k}\in {{\mathbb{C}}^{MN\times {1}  }},
\end{equation}
with
\begin{equation}
\label{equ_channelM}
{{\bf{\Lambda }}_{{k}}} = {\rm{diag}}({[{\lambda _{k,1}} \cdots {\lambda _{k,M}}]^{\rm{T}}}) \otimes {{\mathbf{I}}_N}\in {{\mathbb{C}}^{MN\times {MN}  }}
\end{equation}
where ${\lambda _{km}}\buildrel \Delta \over = d_{k,m}^{ - \zeta }{s_{k,m}}$ represents the large-scale fading between user $k$ and RAU $m$. ${d_{k,m}}$ is the distance from user $k$ to RAU $m$; $\zeta $ is the path loss exponent; ${s_{k,m}}$ is a log-normal shadow fading variable. In addition, ${{\mathbf{h}}_k}\in {{\mathbb{C}}^{MN\times {1}  }}$ models small-scale fast fading, which follows ${{\cal{N}}_{\rm{c}}}\left( {0,1} \right)$. As widely recognized in most of literature, we also assume that the large-scale fading ${\bm{\Lambda }}_{k,m}$ is known to RAU $m$, and the small-scale fading ${{\mathbf{h}}_{k,m}}$ needs to be estimated at RAU $m$.

\subsection{Uplink Channel Estimation}
It is assumed that there are ${\tau}$ orthogonal pilot sequences to be used for uplink channel estimation with ${\tau}<K$, which are denoted by the mutually orthonormal basis vectors $\left\{ {{{\bf{s}}_1} \cdots {{\bf{s}}_{\tau}}} \right\}$, where ${{\bf{s}}_b}\in\mathbb{C}^{\tau}$ is a vector whose $b{\rm{th}}$ element is one, while all other elements are zero. Using the pilot design in \cite{TVan}, the pilot signal ${{\bm{\phi}}_{k}}$ of user $k$ can span
arbitrarily over the above ${\tau}$ basis vectors with different power coefficients, which can be shown as
\begin{equation}
{{\bm{\phi}}_{k}}\in {{\mathbb{C}}^{\tau \times 1}}=\sum\limits_{b = 1}^\tau\sqrt{p_k^b}{{\bf{s}}_b}, \forall k,
\end{equation}
where ${p_k^b}$ is the power that user $k$ assigns to the $b{\rm{th}}$ pilot basis.
The received signal ${\bf{Y}}\in\mathbb{C}^{MN\times \tau}$ at all RAUs is given by
\begin{equation}
\label{equ_y}
{\bf{Y}} = \sum\nolimits_{j = 1}^K {{{\bf{g}}_j}{\bm{\phi}} _j^{\rm{H}}}  + {\bf{N}},
\end{equation}
where ${\mathbf{N}}\in {{\mathbb{C}}^{MN\times \tau }}$ is spatially independent additive white gaussian noise (AWGN) with zero mean and variance $\sigma _n^2$.

Correlating ${{\bf{Y}}}$ in (\ref{equ_y}) with the pilot signal ${{\bm{\phi}}_{k}}$ of user $k$, we can obtain
\begin{equation}
\label{equ_LS}
{\bf{y}}_{k}{\rm{ = }}{\bf{Y}}{{\bm{\phi}}_{k}}= \sum\limits_{j = 1}^K \sum\limits_{b = 1}^{\tau}{\sqrt{{p_k^b}{p_j^b}}} {{\mathbf{g}}_j} + {{\mathbf{N}{{\bm{\phi}}_k}}}.
\end{equation}
Considering minimum mean-square error (MMSE) channel estimator, we have
\begin{equation}
\label{equ_MMSE}
\widehat {\bf{g}}_k{\rm{           =  }}\sum\limits_{b = 1}^{\tau}{p_k^b}{{\bf{\Lambda}}_k}{{\bf{Q}}_k^{ - 1}}{\bf{y}}_{k},
\end{equation}
where
\begin{equation}
{\bf{Q}}_k ={\sum\limits_{j = 1}^K {{{\left( \sum\limits_{b=1}^{\tau }{\sqrt{p_{k}^{b}p_{j}^{b}}} \right)}^{2}}{{\bf{\Lambda}}_j} + {{\sigma _n^2}}\sum\limits_{b=1}^{\tau }{p_{k}^{b}}{{\mathbf{I}}_{MN}}} }
\end{equation}

Furthermore, the mean square error (MSE) of $\widehat {\bf{g}}_k$ is ${\rm{MSE}}_k=\sum\limits_{m = 1}^M \pi_{k,m}$, where
\begin{equation}
\label{equ_mse_PC}
\pi_{k,m}{\rm{ = }}\frac{N{{\lambda }_{k,m}}\left( \sum\limits_{j\ne k}{{{\left( \sum\limits_{b=1}^{\tau }{\sqrt{p_{k}^{b}p_{j}^{b}}} \right)}^{2}}{{\lambda }_{j,m}}}+\sigma _{n}^{2}\sum\limits_{b=1}^{\tau }{p_{k}^{b}} \right)}{\sum\limits_{j=1}^{K}{{{\left( \sum\limits_{b=1}^{\tau }{\sqrt{p_{k}^{b}p_{j}^{b}}} \right)}^{2}}}{{\lambda }_{j,m}}+\sigma _{n}^{2}\sum\limits_{b=1}^{\tau }{p_{k}^{b}}}.
\end{equation}

\subsection{Proposed Problem}
In this paper, we want to optimize the pilot power allocation ${\left\{p_k^b\right\}}$ to minimize the sum MSE of channel estimation. With the assumption that the total pilot transmit power for each user is a predefined constant, the sum MSE minimization problem can be stated as
\begin{alignat}{1}
\setcounter{equation}{9}
\mathcal{P}_0: \min_{\left\{p_k^b\right\}} & \sum\limits_{k=1}^{K}\sum\limits_{m=1}^{M}\pi_{k,m} \tag{9a}\\
    \mathrm{s.t.} & \sum\limits_{b = 1}^{\tau}{p_k^b} = {p_k^{\rm{tot}}},\forall k, \tag{9b}
\end{alignat}
where ${p_k^{\rm{tot}}}$ is the total pilot transmit power for user $k$. The considered problem is NP-hard because of the nonconvexity of the objective. In next section, we propose to design a DNN to solve the problem in an end-to-end fashion. In the proposed scheme, the nonlinear mapping from the channel large-scale fading coefficients $\left\{\lambda_{k,m}\right\}$ to the optimal pilot power allocation ${\left\{p_k^b\right\}}$ is learned by training the DNN with the sum MSE serving as the loss function.
\begin{rem}
Tradition pilot assignment problem assumes that the total pilot transmit power for each user is fully assigned to one pilot sequence, which is a special case of the considered problem $\mathcal{P}_0$.
\end{rem}

\section{Unsupervised Learning Based Pilot Power Allocation}

\subsection{Network Design}
\begin{figure}
  \centering
  \includegraphics[width=8cm,height=4cm]{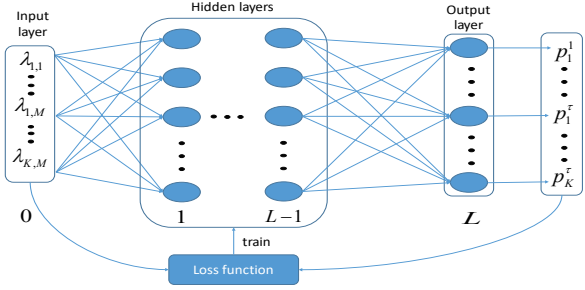}\\
  \caption{The structure of the unsupervised learning based DNN}\label{fig_Illu}
\end{figure}

A fully connected DNN is exploited to address the pilot power allocation problem $\mathcal{P}_0$. The network structure is illustrated in Fig. \ref{fig_Illu}. Specifically, the network consists of one input layer of $KM$ nodes, one output layer of $K{\tau}$ nodes and $L-1$ fully connected hidden layers. Let $\mathcal{L}=\left\{0, \cdots, L\right\}$ represent the set of layers, where $l=0$ and $l=L$ denote the input layer and output layer respectively. The number of nodes of each layer $l\in\mathcal{L}$ is denoted by $n_l$, and we have $n_0=KM$ and $n_L=K{\tau}$.

\begin{figure}
  \centering
  \includegraphics[width=8cm,height=4cm]{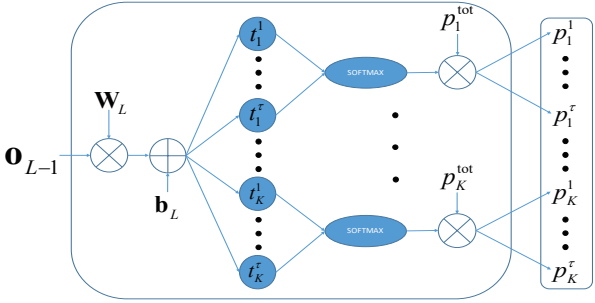}\\
  \caption{The detailed structure of the output layer}\label{fig_outputl}
\end{figure}

The input of the network is formed by aligning the channel large-scale fading coefficients $\left\{\lambda_{k,m}\right\}$ as a column vector, denoted as ${\bf{q}} = {\left[ {{\bf{q}}_1^{\rm{T}}, \cdots ,{\bf{q}}_K^{\rm{T}}} \right]^{\rm{T}}}$, where ${{\bf{q}}_k} = {\left[ {\lambda_{k,1}, \cdots ,\lambda_{k,M} } \right]^{\rm{T}}}$. The output of the network is the pilot power allocation vector ${\bf{p}} = {\left[ {{\bf{p}}_1^{\rm{T}}, \cdots ,{\bf{p}}_K^{\rm{T}}} \right]^{\rm{T}}}$, where ${{\bf{p}}_k} = {\left[ {p_k^1, \cdots ,p_k^\tau } \right]^{\rm{T}}}$. For each hidden layer $l$, the output ${\bf{o}}_l\in {{\mathbb{R}}^{n_l\times 1}}$ is calculated as follows
\begin{equation}
{{\bf{o}}_l} = \left\{ \begin{aligned}
&{\rm{ReLU}}\left( {{\rm{BN}}\left( {{{\bf{W}}_l}{\bf{q}}} \right)} \right), &&l = 1,\\
&{\rm{ReLU}}\left( {{\rm{BN}}\left( {{{\bf{W}}_l}{{\bf{o}}_{l - 1}} + {{\bf{b}}_l}} \right)} \right), &&l \in \left\{ {2, \cdots ,L - 1} \right\},
\end{aligned} \right.
\end{equation}
where ${{\bf{o}}_{l - 1}}\in {{\mathbb{R}}^{n_{l-1}\times 1}}$ is the output of the $\left(l-1\right){\rm{th}}$ layer; ${{\bf{W}}_l}\in {{\mathbb{R}}^{n_{l}\times n_{l-1}}}$ and ${{\bf{b}}_l}\in {{\mathbb{R}}^{n_{l}\times 1}}$ are respectively the weight matrix and bias vector at layer $l$. Note that because the input large-scale fading coefficients $\left\{\lambda_{k,m}\right\}$ are very small, the bias vector in the hidden layer 1 is set to $\bf{0}$. Otherwise, the difference between the large-scale fading of different users may be concealed by the bias vector during the training process, which will impair the performance of the DNN. ${\rm{ReLU}}\left(  x  \right)= {\rm{max}}\left(x,0\right)$ is the Rectified Linear Unit function, which introduces nonlinearity to the network; ${\rm{BN}}\left(  \cdot  \right)$ denotes the batch normalization, which improves the performance and stability of the network.

The detailed structure of the output layer is depicted in Fig. \ref{fig_outputl}. The output ${\bf{o}}_{L-1}$ of the hidden layer $L-1$ is processed by a linear transformation and outputs a temporary vector ${\bf{t}}={{{\bf{W}}_L}{{\bf{o}}_{L - 1}} + {{\bf{b}}_L}}$. Then ${\bf{t}} = {\left[ {{\bf{t}}_1^{\rm{T}}, \cdots ,{\bf{t}}_K^{\rm{T}}} \right]^{\rm{T}}}$ is divided into $K$ groups. Each group ${{\bf{t}}_k} = {\left[ {t_k^1, \cdots ,t_k^\tau } \right]^{\rm{T}}}$ is first input to a softmax function $\sigma$, then multiplies a scale factor ${p_k^{\rm{tot}}}$, and eventually outputs the pilot power allocation vector ${{\bf{p}}_k} = {\left[ {p_k^1, \cdots ,p_k^\tau } \right]^{\rm{T}}}$, which can be expressed by
\begin{equation}
\label{equ_outl}
{{\bf{p}}_k} = p_k^{{\rm{tot}}}\sigma \left( {{{\bf{t}}_k}} \right),
\end{equation}
where the $j{\rm{th}}$ element of $\sigma \left( {{{\bf{t}}_k}} \right)$ is calculated by
\begin{equation}
\sigma {\left( {{{\bf{t}}_k}} \right)_j} = \frac{{{e^{t_k^j}}}}{{\sum\nolimits_{b = 1}^\tau  {{e^{t_k^b}}} }},j = 1, \cdots ,\tau.
\end{equation}
It can be easily proved that $\sum\limits_{j = 1}^{\tau}\sigma {\left( {{{\bf{t}}_k}} \right)_j}=1$. According to (\ref{equ_outl}), we have $\sum\limits_{b = 1}^{\tau}{p_k^b} = {p_k^{\rm{tot}}}$, which is just the constraint (7b) in $\mathcal{P}_0$.

\subsection{Network Training}
As mentioned before, \cite{KKim} designed a supervised learning based DNN to provide a near-optimal pilot assignment with very low computational time.
Similarly, if we were able to obtain the global optimal pilot power allocation $\bf{p}$ for a given channel parameters $\bf{q}$, we would design a supervised learning scheme to train the DNN by using the optimal solution as the ground truth. However, the complexity of obtaining the optimal solution increases exponentially with the number of users, which motivates us to train the DNN by unsupervised learning, which can learn from datasets consisting of input data without labeled responses.

Different from traditional unsupervised learning used for clustering or dimensionality reduction, we develop it in a novel fashion. The considered problem $\mathcal{P}_0$ aims to minimize the sum MSE of channel estimation, thus we directly apply the objective in $\mathcal{P}_0$ as the loss function, which is defined as
\begin{equation}
\label{equ_loss}
loss = \frac{1}{{\left| {{{\mathcal{S}}}} \right|}}\sum\limits_{{\bf{q}\in\mathcal{S}}} \sum\limits_{k=1}^{K}\sum\limits_{m=1}^{M}\pi_{k,m},
\end{equation}
where $\mathcal{S}$ denotes the mini-batch of channel large-scale fading coefficients, and ${\left| {{{\mathcal{S}}}} \right|}$ is the number of samples in $\mathcal{S}$. It can be seen from (\ref{equ_mse_PC}) that $\pi_{k,m}$ is a function of pilot power allocation vector $\bf{p}$, which depends on the specific network parameters ${{\bf{W}}_l}, \forall l \in \mathcal{L}$ and ${{\bf{b}}_l}, \forall l \in \mathcal{L}$. Obviously, the loss function in (\ref{equ_loss}) is differentiable with respect to all network parameters. Hence, we design an iterative algorithm to train the DNN, which is similar to the training mechanism in supervised learning. Specifically, in each iteration, we successively utilize backward propagation algorithm to compute the gradient of loss function with respect to all network parameters and then update the values of them with gradient descend algorithm. The output $\bf{p}$ will correspondingly change, further reducing the value of loss function. The above training process continually adjusts all network parameters to minimize the value of loss function until convergence.

\subsection{Complexity Analysis}
The complexity of training phase consists of forward propagation and backward propagation, both of which mainly depend on the complexity of matrix multiplication. Hence, the complexity of training phase is $O\left( 2\left(KM{n_1}+K\tau{n_{L - 1}}+{\sum\limits_{l = 2}^{L-1} {{n_{l - 1}}{n_l}} }\right)st \right)$, where $s$ is the size of mini-batch, and $t$ is the number of iterations.

Once the DNN has been trained, it can be used to solve the consider problem $\mathcal{P}_0$ by directly feeding the large-scale fading coefficients to the trained DNN. Hence, the complexity of solving $\mathcal{P}_0$ mainly depends on forward propagation, which is $O\left( KM{n_1}+K\tau{n_{L - 1}}+{\sum\limits_{l = 2}^{L-1} {{n_{l - 1}}{n_l}} } \right)$.

\section{Simulation Results}
We consider a hexagonal cell with radius $r=500m$, and there are $M=4$ RAUs with $N=2$ randomly distributed in the cell. We adopt the channel model in (2) to generate huge volumes of channel large-scale fading coefficients data for training and testing, where the path loss exponent $\zeta $ is 3, and the variance of shadow fading is 6dB. For the sake of simplicity, the total pilot transmit power for all users ${p_k^{\rm{tot}}}$ are 6 Watts, and the noise power is set as $10^{-8}$ Watts.
For comparison, we consider the following three methods.
\begin{itemize}
  \item \textbf{Average Pilot Power Allocation (APPA):} The total pilot transmit power for each user is apportioned equally among $\tau$ orthogonal pilot sequences.
  \item \textbf{Exhaustive Search based Pilot Assignment (ESPA):} The total pilot transmit power for each user is fully assigned to one pilot sequence. Actually, it is just the traditional pilot assignments problem. We exhaustively search all possible pilot assignment schemes to find the optimal one.
  \item \textbf{Random Pilot Assignment (RPA):} Similar to ESPA method, each user can only be assigned one pilot sequence, and we randomly select a pilot assignment scheme.
\end{itemize}

\begin{figure}
  \centering
  \includegraphics[width=9cm]{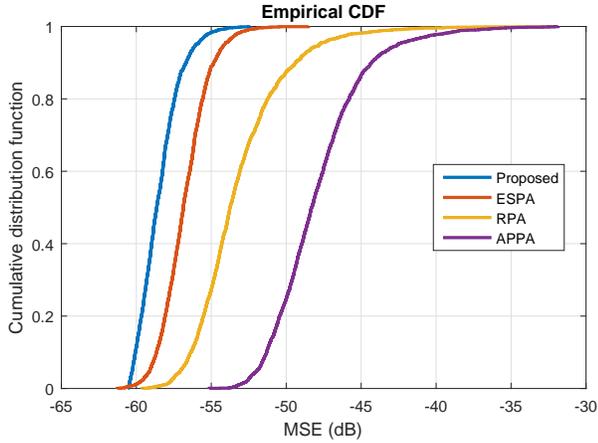}\\
  \caption{The sum MSE performance of the proposed scheme and other methods with $K=12$ and $\tau=4$}\label{fig_mseComp12}
\end{figure}

\begin{figure}
  \centering
  \includegraphics[width=9cm]{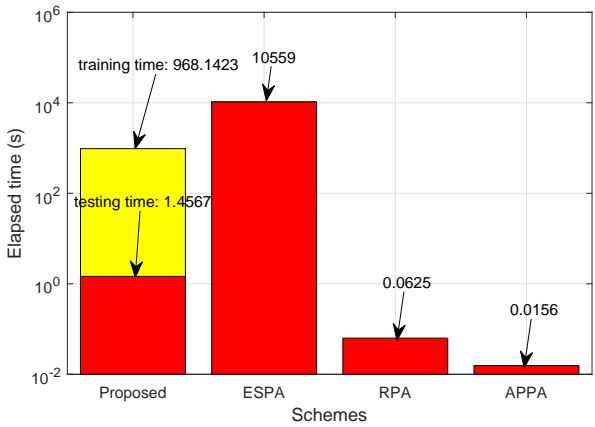}\\
  \caption{The elapsed time of the proposed scheme and other methods with $K=12$ and $\tau=4$}\label{fig_elapsedtime}
\end{figure}

Fig. \ref{fig_mseComp12} and Fig. \ref{fig_elapsedtime} respectively show the sum MSE performance and elapsed time of the proposed scheme and three comparative methods with $K=12$ and $\tau=4$. The input size is $KM=48$, and the output size is $K\tau=48$. The first and the last hidden layer contain 64 neurons, and other three hidden layers have 128 neurons. The mini-batch size $s=1000$, and a total of 1000 iterations are executed to train the network. The cumulative distribution function (CDF) curves in Fig. \ref{fig_mseComp12} and the heights of red bars in Fig. \ref{fig_elapsedtime} are from a testing dataset with 2000 samples. For comparison, all methods are implemented on a computer with intel core i7-8700CPU@3.20GHz and 16GB RAM.

It can be seen from Fig. \ref{fig_mseComp12} that the proposed scheme achieves the best sum MSE performance, followed by ESPA, RPA and APPA. Correspondingly, the elapsed time of all methods are showed in Fig. \ref{fig_elapsedtime}. Here, the RPA and APPA have the lowest elapsed time because of $O\left( 1\right)$ complexity. The ESPA requires 10559s to find the optimal pilot assignment with $O\left( \tau^K\right)$ complexity. The proposed scheme only requires 1.4567s to solve the considered problem $\mathcal{P}_0$ for 2000 channel samples. Although it takes 968.1423s to train the DNN, network training can be performed at a much longer scale than the channel block duration.

\section{Conclusion}
In this letter, we proposed a deep learning based pilot design scheme to alleviate pilot contamination for multi-user distributed massive MIMO systems. By leveraging the unsupervised learning strategy, the trained DNN can be used to solve the sum MSE minimization problem online. Simulation results show that the proposed scheme achieves the best sum MSE performance with low complexity compared to other methods.

\bibliographystyle{IEEEtran}
\bibliography{ps}
\end{document}